# Is Your Data Gone? Comparing Perceived Effectiveness of Thumb Drive Deletion Methods to Actual Effectiveness

Sarah Diesburg, C. Adam Feldhaus, Mojtaba Al Fardan, Jonathan Schlicht, Nigel Ploof
University of Northern Iowa
{sarah.diesburg, adam.feldhaus, alfardm, schlijaa, ploofn}@uni.edu

## ABSTRACT
Previous studies have shown that many users do not use effective data deletion techniques upon sale or surrender of storage devices. A logical assumption is that many users are still confused concerning proper sanitization techniques of devices upon surrender. This paper strives to measure this assumption through a buyback study with a survey component. We recorded participants' thoughts and beliefs concerning deletion, as well as general demographic information, in relation to actual deletion effectiveness on USB thumb drives. Thumb drives were chosen for this study due to their relative low cost, ease of use, and ubiquity. In addition, we also bought used thumb drives from eBay and Amazon Marketplace to use as a comparison to the wider world.

We found that there is no statistically significant difference between buyback and market drives in terms of deletion methods nor presence of sensitive data, and thus our study may be predictive of the perceptions of the market sellers. In our combined data sets, we found over 60% of the drives tested still had recoverable sensitive data, and in the buyback group, we found no correlation between users' perceived versus actual effectiveness of deletion methods. Our results suggest the security community may need to take a different approach to increase the usability, availability, and/or necessity of strong deletion methods.

## Categories and Subject Descriptors
D.4.2 [**Storage Management**]: Secondary storage, D.4.6 [**Security and Protection**], H.1.2 [**User/Machine Systems**]: Human factors

## 1. INTRODUCTION
Personal electronic privacy is increasingly important, and one area of privacy is in the deletion of data on personal storage devices. It is common knowledge in the file systems and storage community that files merely deleted with the delete key or emptied from the trash/recycle bin may be recovered with forensics tools or "undelete" programs. This is because typical file deletion only updates the file's metadata (e.g., pointers to the data) for bookkeeping purposes, while often leaving the file data intact. Even formatting the file system with the default "quick" option does not ensure secure deletion. For example, the MSDOS format command has been shown to only overwrite 0.1% of the data [11]. As a consequence, data thought to have been erased may still be retrieved from a wide array of decommissioned storage devices, including hard drives [11, 42] and USB storage devices like thumb drives [21].

In addition, the growing availability of mobile computing and storage devices encourages sensitive data to be portably stored. For example, an employee can carry a laptop that holds thousands of Social Security numbers, medical histories, and other confidential information. According to a Cisco survey [5], almost one in four employees said they carry corporate data on portable storage devices outside the office.

One answer to this problem is secure deletion, which is concerned with rendering a file's removed content and metadata (e.g., information such as a file's name, size, and owner) irrecoverable. Guidelines have been posted on proper sanitization of electronic storage [14, 15, 26, 40], and many business and organizational policies require data to be erased from storage once it is no longer needed [16, 13, 35, 10]. Also, many free secure deletion tools have been made widely available to the public.

With all of the guidelines, information, and tools available concerning the sanitization of storage media, many users still do not use effective secure deletion methods to sanitize storage devices. A popular assumption is that many users are confused or unknowledgeable concerning best practices of sanitization of data from their devices upon surrender. Is this really true? To test this assumption, we ask the following questions in this study:

1. Do users that have more potentially-sensitive data on their device use better deletion methods?

2. Does any correlation exist between users' perceived versus actual effectiveness of deletion? In other words, do users that use weak deletion methods understand that the method is weak, and vice-versa?

3. What types of potentially-sensitive data do users fail to delete?

4. Do the answers to the above questions differ across general demographics information, such as age, gender, and profession/field of study?

We chose to conduct our study using USB thumb drives. Even though their raw numbers may be declining due to more readily-available cloud storage, they are still very common due to their ease of use, ease of sharing, and situations where transferring data over a network may be impractical or too slow. Used thumb drives are also readily sold over the Internet at a comparably cheaper cost than hard drives or solid state storage drives. Even though they may store sensitive data, thumb drives are often treated in a less-secure manner than hard drives (e.g., left on a desk in plain sight, allowed to travel outside organizations with the user, or lost [1, 27]). In short, thumb drives are an often overlooked, yet fruitful, target for those interested in identify theft [28] or gaining other sensitive information such as trade secrets.

To measure users' perceived effectiveness of deletion, as well as other beliefs concerning deletion, we conducted a buyback study in which we asked participants to trade an old thumb drive for a new 8GB USB 3.0 thumb drive while answering questions to a short survey. Our study collected and analyzed 140 participant drives. No identifying information about the participants was recorded, and the study was approved through our institution through IRB #14-0240. In addition, we did not answer participants' questions about if or how they should delete data from their drives.

To supplement our study, we also purchased 120 used thumb drives from eBay and Amazon Marketplace (henceforth called market drives). We purchased drives from a wide variety of sellers in hopes of increasing drive diversity.

In both groups of thumb drives, we analyzed the deletion method used (if any) and categorized the types of potentially-sensitive data found on the drives through inspecting what files are available upon plugging in the drive without any special tools, and by using forensic tools to recover any data deleted in an insecure manner.

The rest of the paper is structured as follows. Section 2 discusses related work. Section 3 gives background on thumb drive storage characteristics, deletion methods, and categories of sensitive data. Section 4 introduces our study methodology. Section 5 gives study results, special cases, limitations, and delimitations. Section 6 gives discussion and future work, and Section 7 concludes.

## 2. RELATED WORK

To the best of our knowledge, our study is the only study that measures drive owners' thoughts and beliefs about deletion and deletion effectiveness through human subjects research. Thus, the majority of our results section is novel and focuses on data gathered from the buyback participant survey in relation to secure deletion methods, beliefs, and effectiveness. However, other studies categorized data found and deletion method from storage devices purchased from secondary markets, and we summarize their contributions here.

Garfinkel and Shelat [11] acquired 158 hard drives on the secondary market between 2000–2002. Forensic tools were used to recover files, and the types of files were categorized. The authors found only 12% of their purchased drives were sanitized using a strong method that overwrites all hard drive sectors with zeros. They found most drives contained data of some kind, and 42 drives had possible credit card numbers.

In 2007, Valli and Woodward [42] investigated 84 hard drives originating from large corporations and found that 23 of the drives yielded significant exposure of sensitive data. A number of special interest cases are identified in which disks contained illicit, personal, and corporate-identifiable information.

Jones et al. [20] obtained 346 used disks spanning five countries over 5 years (2005–2009). The authors found a consistent improvement in the reduction of both commercial and personal data found over the study period, with the lowest level falling to 30% of drives containing individual data.

Jones et al. [21] also performed a study in which they purchased and analyzed 43 USB storage devices bought from auctions, fairs, and eBay. The study found that 95% of drives contained data that could easily be recovered, 4% of the drives had been effectively sanitized, and 49% of the readable drives had been deleted or formatted, but still contained recoverable data.

In 2011, Sansurooah and Szewczyk [34] acquired, forensically imaged, and analyzed 80 used USB storage devices from eBay Australia. They found 87% of drives did or contained attempts to delete data, and, out of those, 4% of drives were completely sanitized of all data. 95% of the drives contained recoverable data, and 13% of drives showed no evidence of any deletion attempts.

Table 1 illustrates key differences between each study. Note that data classifications are often ambiguous (e.g. what exactly constitutes personal data?) and cannot be directly compared across studies.

**Table 1. Comparison of related work.**

| Study | Storage device | Deletion methods identified | Data classifications | Human subjects study |
|---|---|---|---|---|
| [11] | Hard drives | Some* | File types, CCNs, mail headers | No |
| [42] | Hard drives | Some* | Personal, commercial, illicit | No |
| [20] | Hard drives | Some* | Individual, commercial, illicit | No |
| [21] | Thumb drives+ | Some* | Individual, commercial | No |
| [34] | Thumb drives+ | Some* | Individual, commercial, file types | No |
| This study | Thumb drives | Full** | Personally-identifiable, corporate commercial, corporate confidential, illicit | Yes |

+ Study uses term "USB Storage", but thumb drives are inferred
* Does not measure all possible deletion methods described in Section 3.2
** Measures all possible deletion methods described in Section 3.2

## 3. THUMB DRIVES AND DELETION TECHNIQUES

Even though both are common storage devices, thumb drives are fundamentally different from hard drives. This section gives a brief background on thumb drives, discusses four deletion methods we categorized in our study, and introduces the categories of potentially-sensitive data we searched for on the drives.

### 3.1 USB Thumb Drive Background

The drives collected in our research were all USB thumb drives consisting of a USB connector, one or more NAND flash chips (typically in a TSOP48 package), and a microcontroller device to coordinate access to the NAND flash storage.

The NAND flash chip is a type of memory storage. In addition to reads and writes, it also accepts erase requests at the raw level. However, only the typical read/write hard drive interface is exported to the operating system.

NAND flash must implement specialized remapping and wear-leveling algorithms due to its physical characteristics: (1) In-place updates are generally not allowed—once a location is written, the location must be erased before it can be written again; (2) NAND reads and writes occur to smaller flash pages, but erasures are performed on larger flash blocks consisting of contiguous pages; thus, before a written page can be written again, the flash block containing this page must be erased, and other in-use pages in the same flash block need to be copied elsewhere; and (3) each storage location can be erased only 10K-1M times [7].

These physical challenges are commonly solved through the addition of a flash translation layer (FTL). The FTL sits atop the raw NAND flash interface, typically in the hardware. For backward compatibility, the FTL exports the hard drive interface that accepts only read and write requests; thus, the flash blocks cannot be erased directly[1]. As a common optimization to mask slow writes and erases, when flash receives a request to overwrite a flash page, the FTL remaps the write to a pre-erased flash page, stamps the page with a version number, and marks the old page as invalid[2] to be cleaned later during the garbage collection cycle. These invalid pages are not accessible to components above the block layer, but can be recovered by forensic techniques [4]. To prolong the lifespan of the flash, wear-leveling techniques are often used to spread the number of erasures evenly across all storage locations.

We investigate data which may be retrieved through the microcontroller interface and high-level software forensics tools. In other words, the threat model for this research concerns an attacker using software tools only. Specialized hardware forensic attacks that look at the raw NAND flash chip [4] are beyond the scope of this work and are future work.

Most thumb drives come pre-installed with a variant of the FAT file system [25]. If a drive was found to have an intact non-FAT file system, that drive was excluded from our analysis. Drives were also excluded if files were not in the English language or the drives did not work. Analyzing drives formatted in other file systems or with files in non-English languages is future work. We also attempted to purchase drives of size 8GB or smaller to speed analysis of data on the drives. The breakdown of drives excluded from our study is as follows. See Table 2.

**Table 2. Breakdown of drives eliminated from the study.**

|  | Buyback Drives | Market Drives |
|---|---|---|
| Initial total | 140 | 120 |
| Exclusion: non-FAT file system | 4 | 1 |
| Exclusion: non-English language | 4 | 0 |
| Exclusion: did not work | 6 | 2 |
| **Final study totals** | **126** | **117** |

## 3.2 Deletion Methods

We focus our study on the deletion actions taken before the drive is sold or traded. Thus, if the drive shows evidence of past file deletions, but not all files have been deleted, we categorize the drive as not having been deleted. Similarly, we do not attempt to determine the amount of times in the past a drive has been formatted if it currently contains a live file system.

In order to determine the method used for deletion, binary images of flash memory were inspected using a popular binary file editor called hexedit [17]. We categorized the drives in the following ways:

- *Not deleted (none):* The drive contains files readable by the operating system when plugged into a computer. Any computer user could see these files without any special tools.

- *Normal delete:* The drive does not contain readable files when plugged into a computer, but the drive shows evidence of deleted files that were deleted by 1) emptying a trash can, 2) selecting the file and pressing the "delete" key, or 3) using a command line delete or remove command standard to the operating system (such as Linux rm or Windows del). This type of deletion is evident in the binary image by observing that the first byte of a file or directory's name in a directory entry has been changed to the special hex number 'E5'. Files deleted using a normal delete may be recoverable using high-level software forensics tools.

- *Quick format:* The drive shows evidence of a quick format, which can be performed in Windows by using the format command with the /q flag or by right-clicking on the drive and selecting format with the "quick format" option checked. In Linux, a quick format can be performed with the mkdosfs command, and in Mac OSX by using the Disk Utility with the defaults enabled. A quick format can be identified by the absence (or zeroing) of the file allocation table, which is an indexing structure in the FAT file system. The user has performed a quick format if the file allocation table is zeroed but old entries exist in the FAT file system data section. Files deleted using a quick format may be recoverable with high-level software forensic tools.

- *Full format:* A full format had occurred when the FAT file system data and file allocation tables have been overwritten by zeros, and a clean FAT file system is present. A full format may be performed in Windows by using the format command without the /q flag or by right-clicking on the drive and selecting format with the "quick format" option unchecked. Files deleted using full format are not recoverable with high-level software forensics tools.

- *Other:* Methods that did not fall into the previously discussed categories were classified as other. This category includes methods such as overwriting the drive with random characters or overwriting with single/alternating characters[3]. No readable file system is present. Files deleted using other methods are not recoverable with high-level software forensics tools.

## 3.3 Categories of Data

Files on each thumb drive were classified according to four types of potentially-sensitive data. We expand the definitions of Valli and Woodward [42] to include four categories discussed below:

- *Personally Identifiable Information (PII)*: We used the standard set by the National Institute of Standards and Technology [24], which includes full names, addresses, SSNs, IDs, credit card numbers, and faces.

- *Corporate Commercial (CComm)*: Any piece of information about a company or organization that could, but not necessarily, be made publicly available, such as advertising, product manuals, and product specifications.

- *Corporate Confidential (CConf)*: Any piece of information that originated from a company or organization that which

---

[1] The TRIM command [36] is implemented to optimize performance and is not guaranteed to erase pages/blocks [22].

[2] Flash overwrites might be allowed for some special cases, such as marking a page invalid.

[3] Encrypted devices may appear random [39] and therefore indistinguishable from a random overwrite. Thus, we are not able to definitively tell if these devices are encrypted or randomly overwritten. Either way, no data is recovered.

the company or organization would not want publicly available or could be illegal to release. This includes lists of customers/employees that contained PII, trade secrets, and/or sales/marketing plans.

- *Illicit:* Anything that is pornographic or illegal in nature. By illegal in nature, we mean data that could serve as evidence of potential crimes.

On occasion some pieces of data found fell into two or more categories. We counted these cases in every category applicable.

It was beyond the scope of the research to determine if music, movies, and TV shows were downloaded legally or illegally; thus, we did not include those file types in our potentially-sensitive categorizations. More discussion on this topic can be found in Section 6.

## 4. METHODOLOGY

This section is partitioned into three subsections. The first subsection gives background on how the market drives were obtained. The second subsection discusses the content of the survey presented to participants in the thumb drive buyback program, and the third subsection gives details on the procedures used in processing and categorizing all thumb drives.

### 4.1 Market Drives

We purchased 87 used thumb drives from eBay and 33 used thumb drives from Amazon. In both market purchases, our primary differentiating factor was cost (cheapness) and small size of drive (mostly under 8GB).

The eBay thumb drives were purchased between May and June of 2014 in mostly small lots of 1–2 drives per purchase. Three purchases were made in lot sizes of 8–10, but the lots showed diversity in thumb drive brands and sizes.

The Amazon thumb drives were purchased between March and May of 2014. The majority of thumb drives were single purchases from multiple sellers, with the largest lot group from a single seller being of size three.

### 4.2 Buyback Survey

We recruited participants for the buyback study through flyer distribution on a university campus and email advertisements to university students and employees. The study was conducted inside a large Midwestern university and was open to participants in and outside the university aged 18 or over.

No directly personal identifiable data was collected or used as an identifier. Instead, an alphanumeric identifier was assigned to all collected thumb drives and surveys. Any incidental data found on thumb drives that could possibly identify a participant was kept on a restricted university data server accessible only from researcher computers connected via IPsec, and the thumb drives themselves were kept in a locked university room. After expiration of the IRB, all thumb drive data images will be securely deleted, and all participant thumb drives will be physically destroyed. All researchers and survey administrators have completed a CITI class for training in human subjects research prior to starting this research [6]. In addition, no deception was used in our advertising or recruitment materials. Our survey consent form states: "This study involves research in order to determine user habits and expectations of data deletion as well as if data remains on USB drives despite user efforts to remove data."

All buyback survey administrators were instructed to not provide answers to any types of the following questions concerning:
- If the participant needs to erase his or her drive.
- How the participant should erase his or her drive.

This was done to best emulate the actions taken by users that sell or trade their used thumb drives through other means (such as eBay or Amazon).

We recognize this study is not a perfect mirror of users selling or trading used thumb drives through the market, as participants may be more inclined to trust a university-sanctioned study.

The actual buyback survey asked a number of questions to determine users' thoughts and beliefs about deletion (see Appendix). Not all questions were used in this paper.

### 4.3 Drive Analysis Process

All drives obtained from the buyback and the market groups were processed in the same way, and the buyback survey identifiers are tied to individual drives.

The basic drive analysis phase can be broken down into five steps:
1. Record metadata identifying the drive,
2. Create safe and working images for analysis,
3. Classify deletion technique,
4. Classify non-deleted data, and
5. Classify data found with forensics tools.

Each phase is detailed in the following subsections. After each phase, resulting data found was recorded to a database containing results with no personal identifiers. The analysis was performed on Linux Debian 7 workstations.

*Step 1) Identify drive metadata:* The basic characteristics, such as manufacturer, model, identifying marks, and size of the flash drives were identified using the Linux commands '`fdisk -l`' and '`lsusb`'. To assist with cataloging, a photo of each USB flash drive was taken. The information, along with the drive's unique alphanumeric identifier, was entered into a wiki and database for tracking.

*Step 2) Create safe and working drive images:* Two unmodified binary level copies of the drive were taken: a safe copy to be stored in a separate location and a working copy. The safe copy exists to preserve the state of the drive as we received it and was created using the Linux `dd` command. We then copied the safe copy image into a working copy image for which all future forensics analysis steps was applied.

*Step 3) Classify deletion techniques:* We mount the image as a type 'vfat' file system and search for any visible files and directories. If any files or directories were available, we classified the drive as *not deleted* (see Section 3.2) and skip to the next step (classify non-deleted data).

However, if no files or directories are visible, the user has likely deployed some sort of deletion technique. We then use the method described in Section 3.2 to classify the deletion method as a *normal deletion*, *quick format*, *full format*, or *other*.

Through manual analysis of the raw image through the editor `hexedit`, we were able to classify the following deletion techniques used (if any): *normal (or regular) deletion, quick format deletion, full format deletion, or other* deletion methods. We then move to the step 5, classifying data found with forensics tools.

*Step 4) Classify non-deleted data:* The working copy image of each flash drive was mounted and examined for files visible without the use of any forensics tools. Directories were explored both via command line to uncover possible hidden files as well as through a GUI. Files found using normal file browsing methods were grouped into the four different potentially-sensitive categories defined in Section 3.3. Files were then separated into four category-specific directories. When a file qualified for multiple categories, copies of the file were added to additional category directories. The number of bytes found in each category were measured with the Linux command `du -b --apparent-size` on a Debian 7 Linux system to get an exact byte count.

*Step 5) Classify data found with forensics tools:* Each flash drive working image was then analyzed using an open-source forensics tool suite [30, 41]. The files returned were then classified and sorted into the same groups discussed in Step 4. It is important to note that non-deleted files are also "recovered" using the forensic tools, so files discovered in Step 4 are once again counted in this step.

# 5. RESULTS

This section is broken into three subsections. Subsection 5.1 discusses specific examples of sensitive data we were able to recover. Subsection 5.2 compares measurable attributes between the market and buyback drives groups to determine if the buyback participants are representative of market sellers. Finally, subsection 5.3 analyzes buyback participants' thoughts and beliefs concerning deletion compared to their actual deletion methods across age, gender, and profession/field of study.

For statistical purposes in subsections 5.2 and 5.3, the four drives that were deleted using techniques described as 'other' were included in the group 'full format.' We did this because the end result was much the same—a drive with no recoverable data using our techniques. We also used the Pearson's Chi-squared analysis to analyze the relationships among the categorical data. We consistently use α = 0.05 as our significance level for statistical analysis.

## 5.1 Data Recovered

This subsection illustrates the range of sensitive information that can be recovered from second-hand thumb drives from both the buyback and market drive groups.

### 5.1.1 Buyback Drives

The buyback drive marked with identifier B092 (2GB Sandisk drive) had a large variety of sensitive information crossing multiple categories identifying the study participant and the participant's business. The drive included the participant's picture, resume, employment history, family addresses and contact information, a 2011 tax return, medical screening information, a color copy of the participant's passport, military discharge paperwork, a W-4 2012 form, direct deposit information, a copy of a voided check, and lists of names, addresses, phone numbers, and birthdays of employees. The drive also contained commercially produced and homemade pornographic videos. The survey connected to this drive indicated the participant deleted information via the normal deletion method, is ambivalent about deletion (selected "in between" on survey question 5), and thought recoverability of the data would be somewhat easy.

The buyback drive marked with identifier B096 (4GB Sandisk drive) had a recoverable passport. This participant did not delete the thumb drive, yet answered that he or she "always care(s)" about deletion. When asked on the survey why the participant did not delete, the participant responded "in case I needed it for future purposes." This example may be illustrative of the implicit trust of a university study; however, the study description was clear in that the participant's drive would be traded (not given back) and then associated data made anonymous.

The buyback drive marked with identifier B002 (16GB PNY drive) contained a resume, one text file with a Bitcoin wallet private key, and Litecoin wallet encryption and private key. The participant used quick format as a deletion method, "often cares" about deletion, and thought recoverability of the data would be "somewhat easy". When asked about deletion method used, the participant said "I just transferred my flash drive data onto my personal computer." This answer implies the participant did not understand what was meant by the term *deletion*.

### 5.1.2 Market Drives

The market drive marked with identifier M081 (1GB Apacer drive) contained sensitive information about the internal workings of a middle school. The drive contained middle school teachers' pictures, students' pictures, and names of administrators, teachers, and students. It also contained school schedules, plans, announcements, and internal incident report documents. This drive was deleted using a quick format.

The market drive marked with identifier M055 (8GB Sandisk drive) contained on mount a company payroll spreadsheet with personally-identifiable information for multiple employees of the company. The drive also contained invoice information for the company, as well as personal photographs and other papers with individual names that were not related to the company. The drive was deleted using normal deletion.

The market drive marked with identifier M086 (2GB Sandisk drive) contained internal documents for a major clothing retailer in the United States. The drive also contained an individual's tax document with the person's name, address, and social security number. Also present on the drive were personal photos with names and addresses of presumable acquaintances of the previous owner. The drive was deleted using normal deletion.

The market drive marked with identifier M115 (2GB Lexar drive) contained a communication from a state office to a university in that state requesting funding and recognition for a program, as well as several corporate commercial documents pertaining to that university. The drive also contained personal information from an executive-level member of the university—including tax documents, personal photos, and an illicit photo. The drive was deleted using normal deletion.

## 5.2 Comparisons Between Drive Groups

Overall, we found that the drives purchased from the marketplaces were not different from the buyback drives in terms of presence of potentially-sensitive data nor methods of deletion. This lead us to reason that the buyback drives would be similar to the market drives in other ways that were impossible to discern from the buyback drives (i.e. the confidence sellers had in their deletion techniques or the demographic information of the sellers). In that, we found no statistically significant patterns to discern relationships among deletion practices.

First, we compared the market drives to the buyback drives in terms of method of deletion (see Figure 1) and presence of potentially-sensitive data (see Figure 3). Using chi-squared goodness-of-fit analysis, we found the drives not dissimilar to the buyback drives; $X^2$ (3, $N$ = 243) = 1.53, $p$ = 0.33. (As a reminder, we cannot prove similarity of the drive groups, only that the drive groups are not dissimilar.) Of interest, over 33% of drives did not have any attempt to remove data upon surrender.

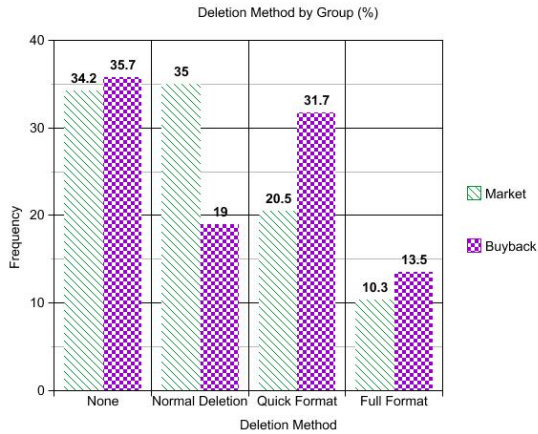

**Figure 1. Deletion method buy group between buyback and market drives as percentages.**

By re-categorizing the deletion methods into methods that yield recoverable data (insecure) versus methods that do not (secure), we can further see the similarities between drive groups. See Figure 2.

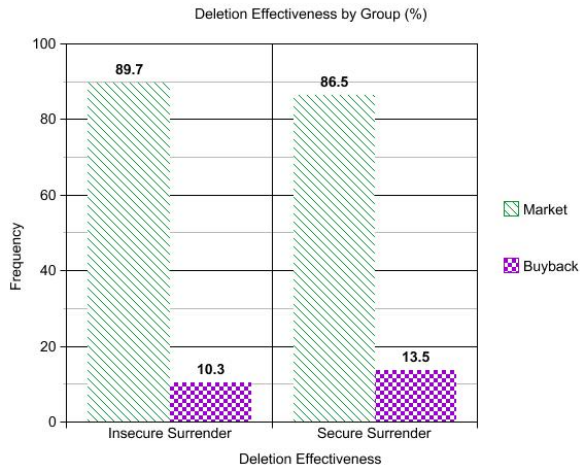

**Figure 2. Deletion effectiveness by group as percentages.**

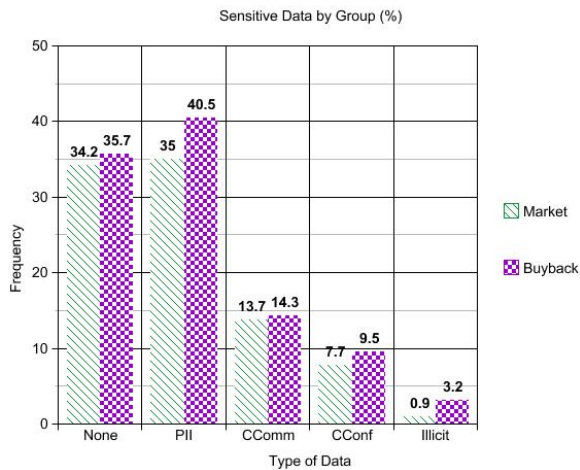

**Figure 3. Sensitive data present on market and buyback drives as percentages.**

Also, we found no statistically significant differences when we compared the market drives to the buyback drives in terms of the presence of potentially-sensitive data. $X^2$ (4, $N$ = 243) = 1.82, $p$ = 0.23 (see Figure 3).

The percentage of potentially-sensitive data found in the market drive group was 65.81%, and the percentage of potentially-sensitive data found in the buyback drive group was 64.29%. In total, 65.02% of data found in both groups was classified as potentially sensitive.

Because of these results, we feel confident that the drive populations may be not dissimilar in other characteristic ways. This means the survey data may be predictive of the general patterns in people surrendering drives.

## 5.3 Buyback Results

We analyzed the drives to compare the presence of potentially-sensitive data with the confidence the participants had in the deletion methods used to clean the drives before surrender. In that, we found no statistically significant relationship between these two variables. $X^2$ (4, $N$ = 90) = 3.17, $p$ = 0.47. See Figure 4. Please note that 36 participants indicated they did not delete any data or did not reply to the question—thus, they were removed from this analysis.

This leads us to conclude that there is no statistically significant relationship between the presence of sensitive data on the drive and the confidence the participant had in removing information from the surrendered drive.

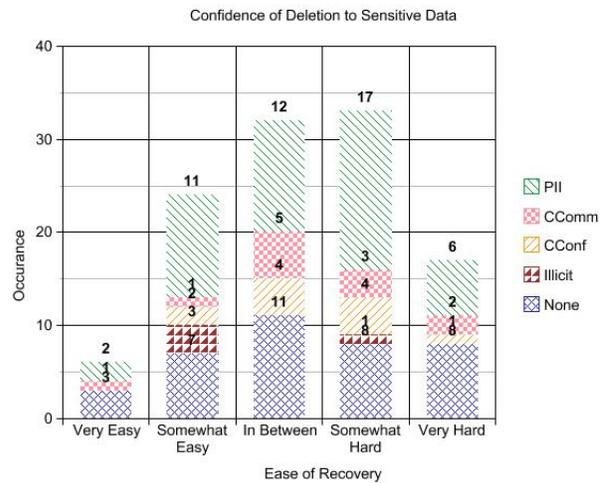

**Figure 4. Presence of sensitive data by perception of data recoverability.**

We also tested to determine if there was a statistically significant relationship between the variables of 'deletion methods' and 'confidence of deletion'. We removed from consideration those who either responded 'did not know' or failed to respond to the question regarding confidence of deletion. As with the other categories, we found no statistically significant relationship ($X^2$ (12, $N$ = 90) = 7.94, $p$ = 0.21). In fact, we see a disturbing uniformity between deletion methods present in all confidence categories (see Figure 5). For instance, we see that 66.7% of those selecting "Very Hard" used methods in which data was recoverable, while in "Very Easy," we found 12.5% of users who did use deletion methods in which data could not be recovered.

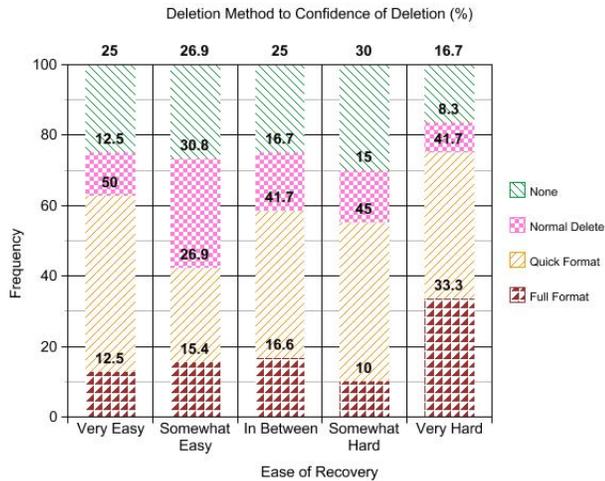

Figure 5. Deletion method compared to confidence of deletion as a percentage.

### 5.3.1 Gender

We had $n = 1$ participant who declined to include their gender, and thus was removed from the proceeding analyses. Of the remaining participants, 44 were female and 81 were male.

We found no statistically significant differences between genders with respect to deletion method ($X^2$ (3, $N = 125$) = 1.37, $p = 0.29$), presence of sensitive data ($X^2$ (1, $N = 125$) = 0.93, $p = 0.24$), nor confidence of deletion ($X^2$ (4, $N = 125$) = 1.92, $p = 0.25$).

There is a slight preference for women to not delete the drive (see Figure 6); however, these values are within the error of the test. Notably, the percentage of men and women that chose the 'full format' (which yields no data) is similar.

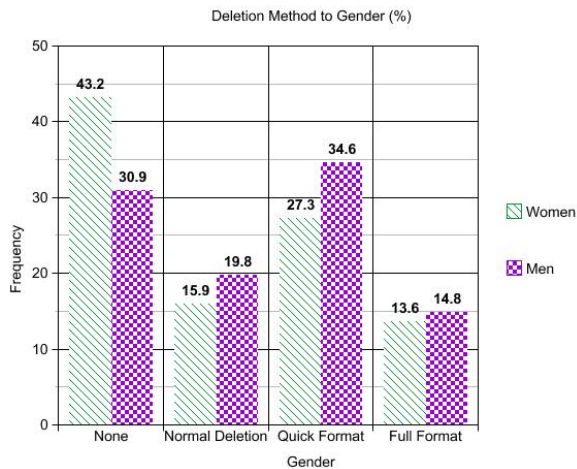

Figure 6. Deletion method by gender as percentages.

We can then re-categorize the deletion methods shown in Figure 6 into insecure surrender and secure surrender methods, similar to what was defined in Figure 2. We find similar percentages of men and women surrendered drives both securely and insecurely. Further, both men and women tend to use insecure methods. See Figure 7. Similarly, we see slight trends for women to have PII contained on their drives (see Figure 8), but these are not statistically significant and do not reflect a general prevalence of potentially-sensitive data. $X^2$ (8, $N = 125$) = 6.95, $p = 0.27$.

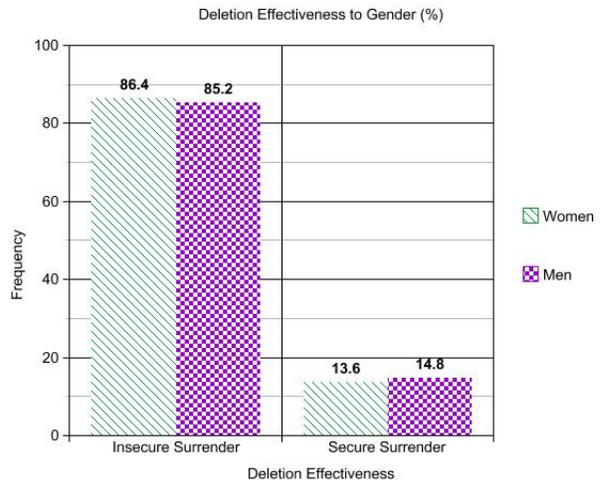

Figure 7. Deletion effectiveness by gender as percentages.

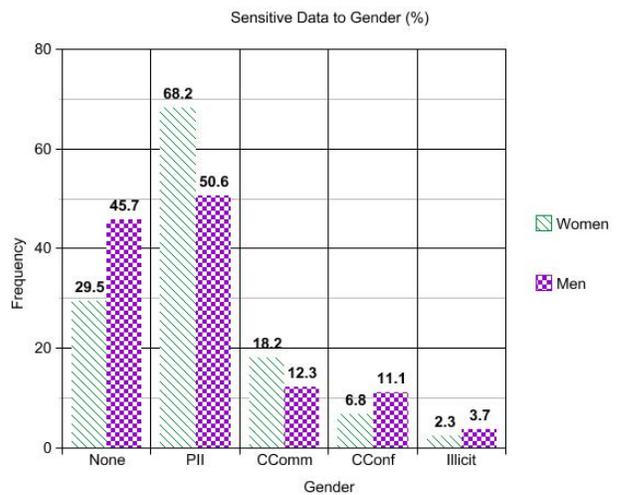

Figure 8. Presence of sensitive data by gender as percentages.

Finally, we see no statistically significant difference in confidence of the participants' deletion methods (see Figure 9).

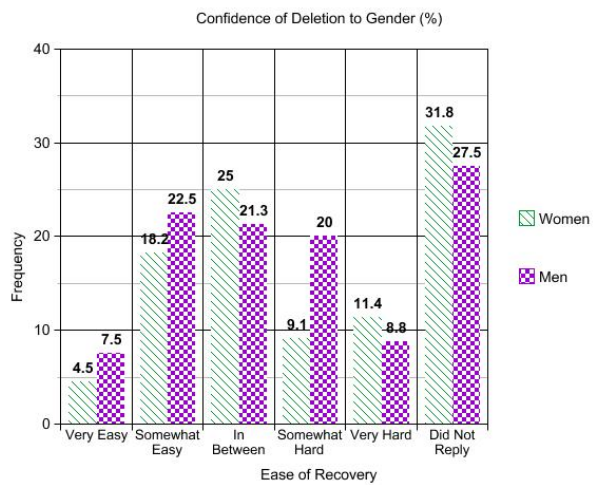

Figure 9. Confidence of deletion by gender as percentages.

This leads us to reason that we cannot discern either the method of deletion, the presence of sensitive data, or the confidence the participant will have in their deletion methods from their gender.

*5.3.2 Age*

To create more even groups in terms of group population while maintaining similar age-bands, we combined the groups from the survey into the following categories: 18–30, 31–40, 41–50, and Over 51. Even with this adjustment, 75/126 = 59.52% of the participants identified themselves as 18–30 (see Table 3).

**Table 3. Total of participants by age bracket.**

| Age Bracket | Total |
|---|---|
| 18 – 30 | 75 |
| 31 – 40 | 19 |
| 41 – 50 | 14 |
| Over 50 | 18 |

As in the previous analysis, we found no statistically significant differences in age with regard to deletion method, ($X^2$ (9, $N$ = 126) = 4.03, $p$ = 0.09, see Figure 10), presence of sensitive data ($X^2$ (3, $N$ = 126) = 1.31, $p$ = 0.27, see Figure 11) nor confidence in deletion methods $X^2$ (12, $N$ = 90) = 3.99, $p$ = 0.09, see Figure 12).

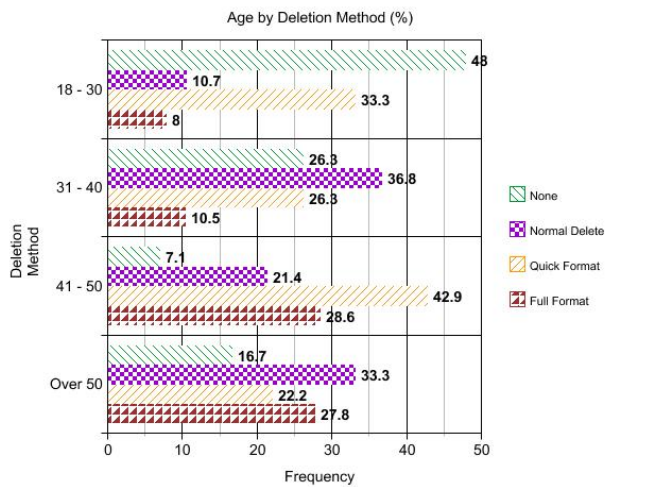

Figure 10. Deletion method with regard to age as percentages.

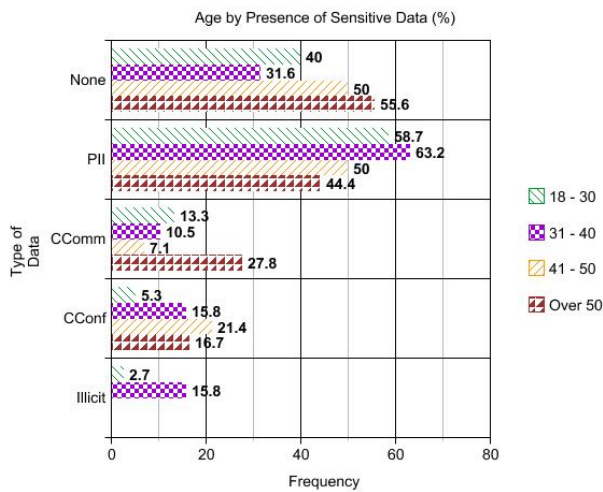

Figure 11. Presence of sensitive data by age as percentages.

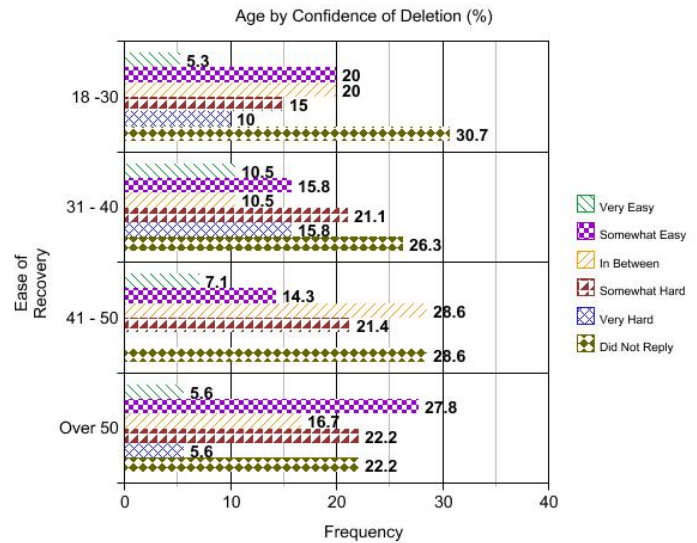

**Figure 12. Confidence of deletion by age as percentages.**

*5.3.3 Profession/Field of Study*

As with gender, we compared the deletion methods, presence of sensitive data, and confidence in deletion method to the pre-determined profession/field-of-study categories. As in the previous section, those who either did not respond to this question or responded 'other' were removed from the analysis. This consisted of $n$ = 4 participants.

It should be noted that we had a majority of participants who self-identified as being in the Science/Engineering category ($n$ = 60, see Table 4). We found no statistically significant relationship between field of study and deletion method ($X^2$ (9, $N$ = 122) = 6.10, $p$ = 0.27, see Figure 13), presence of sensitive data ($X^2$ (9, $N$ = 122) = 5.16, $p$ = 0.18, see Figure 14) nor confidence of deletion $X^2$ (9, $N$ = 126) = 5.27, $p$ = 0.19, see Figure 15).

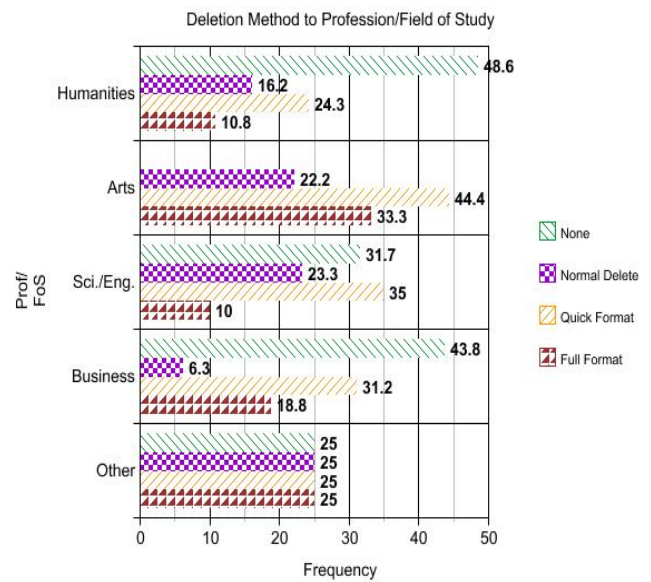

**Figure 13. Deletion method to profession or field of study.**

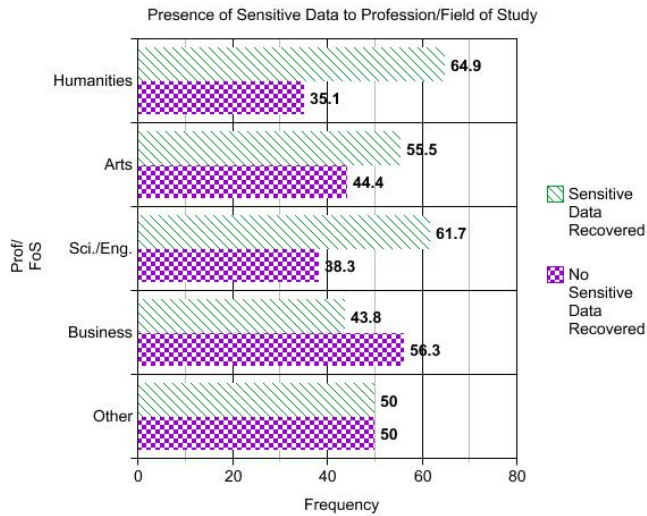

**Figure 14. Presence of sensitive data to profession/field of study.**

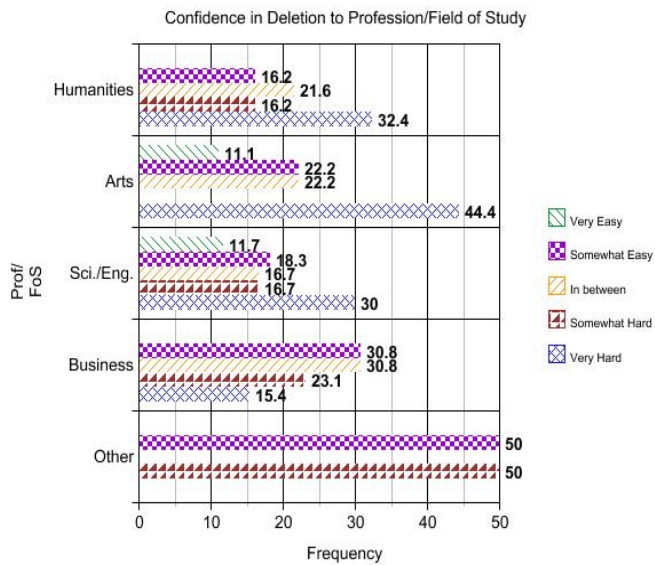

**Figure 15. Confidence of deletion to profession or field of study.**

**Table 4. Total of participants by profession/field of study.**

| Profession/FoS | Total |
|---|---|
| Humanities | 37 |
| Arts | 9 |
| Science/Engineering | 60 |
| Business | 16 |
| Other/Did not respond | 4 |

## 5.4 Limitations and Delimitations

Due to scarcity of resources, we encountered manpower limitations, impacting our ability to parse drives. This in turn affected the number of drives processed. The singular buyback study location, which was physically close to traditionally-STEM departments at a university, skewed the participant field of study. The buyback study could have been held in different and/or multiple locations (such as a mall, public recreation center, business environment, etc.), but we chose to hold it in a university due to resource constraints. Also, conducting the survey in a Midwestern university narrowed our pool of potential buyback participants, as many of the participants were associated with the university.

A delimitation of our study included buying drives from a combination of individual drives and lots of drives. We chose to purchase some lots of drives to help minimize the cost of the study. Those lots demonstrated diversity in drive model, size, and other advertised attributes. We also limited our purchase of market drives to eBay and Amazon Marketplace; however, other avenues to purchase drives may exist. We also chose to use the TestDisk [41] suite of forensics software; however other researchers may have chosen to use different open-source or commercially-available forensics software. We chose this software due to its ease-of-use with the research team.

## 6. DISCUSSION AND FUTURE WORK

The lack of relationship between users' perception of data recovery and method of deletion may arise from any number of factors. We list some possibilities here:

- Lack of awareness in the need to use strong deletion methods.
- Lack of education concerning the use of strong deletion methods.
- Perceived difficulty in using strong deletion methods.

One potential mitigation strategy is to better educate the public on the use of strong deletion methods, such as full formatting or the use of specialized tools such as DBAN. However, this strategy hinges on the users' willingness to learn about additional techniques.

Another strategy is to set strong deletion methods to be the default. One current example of "security by default" is the trend of mobile phone operating systems to encrypt the phone storage by default [2, 3]. If strong deletion were made the default choice, perhaps users would choose strong deletion more often. One example might be setting the full format as the default setting in Windows and OSX, as opposed to the quick format option. Per-file strong (or secure) deletion is more difficult [38, 37, 9, 43, 31, 32, 8, 33], and this option would likely require modifications to the applicable operating system and storage interface.

Even though many organizations include policies for sanitization of removable storage media like thumb drives, thumb drives are often difficult to track. A work thumb drive could easily wind up in an employee's home or in a lost and found, where it may not be properly sanitized upon surrender. As our special cases illustrate, organizations are not immune to this problem. One policy solution is to ban the use of thumb drives all together—however, users may not be willing to give up the convenience a thumb drive provides, and thus the policy may be difficult to enforce.

Physical destruction could properly sanitize the thumb drive, but only if the actual NAND chip itself is destroyed. Phillips et al. [29] discuss interesting ways in which thumb drives can be physically destroyed and whether or not data could be recovered in those instances. The downside to physical destruction is that the drive may never be used again.

Another solution could be to modify the thumb drive to allow some sort of secure deletion button, slider bar, or reset pin. Another technique could be to modify the thumb drive to allow a secure erase command, much like hard drives and solid state drives [18]. These solutions would allow the thumb drive to remain usable after sanitization, but the solutions must be implemented correctly to be of use [43].

Finally, users could 1) use software to encrypt data on their thumb drives or else 2) use self-encrypting thumb drives [19]. The first suggestion is less user-friendly, as the software needed to decrypt the drive must be installed on every computer. The second suggestion is often more expensive. Also, if the key is ever recovered (e.g. through coercion or a brute-force attack on the password used to generate the key), the data on the drive would no longer be deleted.

During sensitive data classification, we found classifying movies, TV shows, and music as illicit (illegal) is difficult for a number of reasons. First, laws and regulations regarding copyrighted materials vary [12, 23]. Secondly, it becomes troublesome to determine if a particular music, TV show, or movie file were actually purchased or archived legitimately. Sometimes, one can infer by the naming or packaging of a file if it likely came from an illegal source, but it is difficult to infer illegality beyond reasonable doubt. Thus, we did not categorize copyrighted works as illicit in this research. Using advanced methods to determine legality of media files will be future work.

Other future work includes expanding the buyback study to different pools of participants in different locations, as well as using hardware forensics methods to remove the NAND chips from the USB drive to read all storage areas on the chip [4].

## 7. CONCLUSION

To understand the deletion beliefs and practices of people surrendering thumb drives in exchange for money or goods, we conducted a buyback program with a survey component and compared the results to drives obtained through common markets such as eBay and Amazon Marketplace. Of interest, we found:

- The percentage of buyback and market drives that contain potentially-sensitive data in our study to be over 60%.
- No statistically significant relationship connecting the presence of potentially-sensitive data and deletion method.
- No correlation between users' perceived versus actual effectiveness of deletion in the buyback group
- Near equal ratios of men and women surrendered drives with recoverable data in the buyback group.
- Used drives in our study contained a variety of sensitive data, including employee databases, tax returns, internal corporate documents, scanned passports, and illicit photos.

We found a general state of confusion about the effectiveness of deletion methods Future work must be performed to remedy the situation, perhaps by better user education, more default secure sanitization policies in operating systems, built-in device mechanisms, or some combination of these solutions.

# APPENDIX

Below is the survey we gave to our buyback participants.

1. Did you delete data from your USB thumb drive?  ☐ Yes  ☐ No

2. (Only answer if you checked no on question 1.)  If you didn't delete data from your USB thumb drive, why not?
_________________________________________________________________________________________________________________

3. (Only answer these questions if you checked yes on question 1.)

   3a. If you deleted your data, please describe the method you used.
   _________________________________________________________________________________________________________

   3b. How easy was it to delete your data?

   ☐ Very Easy   ☐ Somewhat Easy   ☐ In Between   ☐ Somewhat Hard   ☐ Very Hard

   3c. Do you think the deleted data is easy or difficult to recover?

   ☐ Easy to Recover   ☐ Somewhat Easy   ☐ In Between   ☐ Somewhat Hard   ☐ Hard to Recover

4. Estimate how many USB thumb drives you own and use.

   Own? _________________   Use? _________________

5. How much do you care about permanently deleting data on USB thumb drives?

   ☐ Never Care   ☐ Rarely Care   ☐ In Between   ☐ Often Care   ☐ Always Care

6. What is the maximum amount of time you would spend trying to delete your data permanently on a USB thumb drive? _____________________

7. Which category best describes your field of study or work from the following:

☐ Humanities ☐ Arts ☐ Science/Engineering ☐ Business/Management ☐ Other (Please write):

8. What is your gender?

☐ M ☐ F ☐ Trans* ☐ Fill in _____________ ☐ Prefer not to answer

9. What is your age group?

☐ 18-24 ☐ 25-30 ☐ 31-35 ☐ 36-40 ☐ 41-45 ☐ 46-50 ☐ 51-55 ☐ 56-60 ☐ 61-65 ☐ 66+ ☐ Prefer not to answer